# Evaluation to Chinese marine economy in the coastal areas

Yi Zheng

Business School, Shanghai Jian Qiao University, Shanghai 201306，China.
Email: yzheng@shou.edu.cn

**Abstract.** For promoting the development of marine economy more sustainable, based the data envelopment analysis method and combined with the impact of the marine environment, the environmental performance of marine economy was evaluated for Chinese coastal provinces. Firstly the classical CCR model was used. Then a model considered undesirable outputs was made to suit to Chinese marine economy. Made use of the two models, the economic efficiencies without environmental consideration and the environmental performance index were calculated and compared. According to the results, the empirical relation between EPI and EE，per capita GDP and the industrial structure was analyzed. It is useful for guiding the coastal local economy of China to a healthy way.

**Introduction**

With the ocean and coastal resources were developed, marine pollution is becoming more and more serious, it will be the bottleneck for further development of marine economy. So developing in quality and efficiency has become an inevitable requirement for its sustainable development. This requires that the development of marine economy in our country should be properly adjusted. Comprehensive evaluation for marine economic efficiency can find the developing level and existing problem of marine economy in the coastal areas, so as to providing some reference for scientific decision-making and guide the coastal local economy to a healthy way.

In recent years the term "environmental performance", which offers policy analysts and decision makers condensed information by the construction of an aggregated environmental performance index(EPI), has been universally advocated and quoted. From the point of view of operations research, the existing techniques for constructing aggregated EPIs can also be divided into two big categories, namely the indirect approach and direct approach. In the indirect approach, the key economic and environmental sub-indicators are first identified, which are then normalized and integrated into an overall index by some weighting and aggregating techniques. In contrast, in the direct approach, an aggregated EPI is directly obtained from the observed quantities of the inputs and outputs of the studied system using a nonparametric approach called Data envelopment analysis (DEA) [1].

Chung et al．(1997)，Seiford and Zhu(2002)，Vencheh et al．(2005)，Bian Yiwen (2007)，Hua et al．(2008)，Lozano et al．(2009) etc. have researched and developed the problem of environmental efficiency evaluation from different angles with data envelopment analysis. They made a lot of achievements, but the result about the marine industries is almost no in these research. So this paper made use of the DEA method and combined with the impact of the marine environment to evaluate the environmental performance of marine economy for Chinese coastal provinces,. It will promote the development of marine economy more sustainable and also develop the field of the DEA application.

**Research Methods**

**Evaluation Model.** For predictive models to provide reliable guidance in decision making processes, they are often required to be accurate and robust to distribution shifts[2]. It is not easy for our study. So firstly the classical CCR model (output-oriented) in data envelopment analysis [3] was used in this paper. It only considered the desirable production, without considering the harmful effects on the environment. Then another model was made, in which a decision making unit is efficient when it produces the most desirable production with the least damage to the environment. And these byproducts generated with desirable productions, such as sewage, called undesirable output. Obviously, the desirable output and undesirable output always produced at the same time and must be improved in the opposite direction, the more desirable outputs and the least undesirable outputs the better. Although the traditional DEA model have been made some gains in dealing with both the different outputs, but there is still much room for improvement. In order to consider economic and environmental performance better, this paper uses a slack-based measure (SBM) of efficiency in data envelopment analysis to construct a comprehensive efficiency evaluation index by maximizing input and output slack variables.

SBM is an efficiency evaluation model based on DEA theory. It was put forward by Tone (2001) and considered undesirable output with desirable output [4]. It is different from the traditional input - oriented or output-oriented DEA model; SBM puts the two orientations into the same model to improve all possible variables in the objective function. In the end it gets a value between 0 and 1 for efficiency measurement. The SBM efficiency value is not affected by the units of input and output and its values is monotonically decreasing for each input and output slack. So the slack variables were greater, the smaller the SBM efficiency. On this basis, an output-oriented model for undesirable outputs (UOM) was developed to suit to Chinese marine economy as follows.

In a production system with both a desired outputs and undesirable outputs, it is assumed that X is a input matrix, and output matrix Y was decomposed into desired outputs $Y^g$ and undesirable outputs $Y^b$, so the EPI of a decision making unit (DMU) $(x_0, y_0^g, y_0^b)$ was drawn:

$$\rho* = \min \frac{1 - \frac{1}{m}\sum_{i=1}^{m_1}\frac{s_i^-}{x_{i0}}}{1 + \frac{1}{s}\left(\sum_{r=1}^{s_1}\frac{s_r^g}{y_{r0}^g} + \sum_{r=1}^{s_2}\frac{s_r^b}{y_{r0}^b}\right)}$$

subject to  $x_0 = X\lambda + s^-$
$y_0^g = Y\lambda - s^g$
$y_0^b = Y\lambda + s^b$ (1)
$L \leq e\lambda \leq U$
$s^-, s^g, s^b, \lambda \geq 0,$

Where $s_1$, $s_2$ are the number of species in $Y^g$ and $Y^b$; $s = s_1 + s_2$. The DMU $(x_0, y_0^g, y_0^b)$ is efficient if and only if $\rho* = 1$. If the DMU is inefficient, i.e., $\rho* < 1$, it can be improved and become efficient by deleting the excess in inputs and undesirable outputs and augmenting the shortfalls in desirable outputs, the corresponding amount is respectively $s^-$, $s^b$ and $s^g$.

**Index Selection.** To establish a rational evaluation index system is important for efficiency evaluation in data envelopment analysis, because the different index system will get the different evaluation result. So the reasonable selection of the index system is the basis of correct evaluation in this paper.

According to rule of thumb in the DEA theory, DMU's number is generally about two times the index number. The more index number will lead to more effective DMUs, and reduce the discrimination of the model; the less index number is easy to cause the measure results have certain one-sidedness, and can't provide sufficient information to support managers' decision. In this paper, for evaluating marine economy, 11 coastal provinces in Chinese mainland were selected as 11 DMUs, so the index number of input and output is suitable for 5-6. For reflecting the undesired output and the diversity of marine industries as much as possible, the upper limit at 6 indexes was selected, four input indexes were national ocean-related employed personnel, ownership of marine fishing motor vessels, rooms in star-related hotels by coastal regions and berths for productive use at above designed size seaports, a desirable output index was gross ocean product, and an undesirable output index was total volume of industrial waste water discharged.

The coastal tourism industry, marine transportation and marine fishery, which were reflected by the above input indexes, were the top three for added value of marine industry in 2011, and they accounted for respectively 33.08%, 22.36% and 16.98% in total added value [5]. So the above index selection was a relatively rational choice in the condition of the index number must be limited, because they reflect more than 70% of China's marine industry, although can't reflect the total.

**Data and Results**

**Data Collection**. The data needed by indexes is about 11 coastal provinces and municipalities. It is Liaoning, Hebei, Tianjin, Shandong, Jiangsu, Shanghai, Zhejiang, Fujian, Guangdong, Guangxi and Hainan. In them, the data about national ocean-related employed personnel, star-related hotel rooms by coastal regions and berths for productive use at above designed size seaports, gross ocean product, and total volume of industrial waste water discharged in 2011 was collected in "China Marine Statistical Yearbook 2012"; and the ownership of marine fishing motor vessels was come from "China fishery Statistical Yearbook 2012". The statistical properties of the data was showed in table 1, and table 2 gave the correlation matrix between the input and output data.

Table 1  Descriptive statistics on inputs and outputs data in 2011

| inputs and outputs | Personnel (10000persons) | Fishing vessels (unit) | Berths (number) | Number of Hotel Rooms（unit） | Waste Water Discharged （10000t） | Gross Ocean Product （100 million yuan） |
|---|---|---|---|---|---|---|
| Max | 820.4 | 59057 | 1392 | 140252 | 246298.5 | 9191.1 |
| Min | 94.2 | 543 | 53 | 16850 | 6820.12 | 613.8 |
| Average | 309.31 | 26382.3 | 430.27 | 67469.09 | 123002.84 | 4136 |
| SD | 211.05 | 20126.2 | 415.66 | 36289.46 | 74293.07 | 2625.9 |

Table 2  Correlation matrices for inputs and outputs data in 2011

| | Personnel | Fishing vessels | Berths | Number of Hotel Rooms | Waste Water Discharged | Gross Ocean Product |
|---|---|---|---|---|---|---|
| Personnel | 1.00 | 0.79 | 0.77 | 0.81 | 0.51 | 0.86 |
| Fishing vessels | 0.34 | 1.00 | -0.06 | 0.25 | 0.29 | 0.29 |
| Berths | 0.77 | 0.47 | 1.00 | 0.73 | 0.29 | 0.62 |
| Number of Hotel Rooms | 0.81 | 0.50 | 0.73 | 1.00 | 0.75 | 0.79 |
| Waste Water Discharged | 0.51 | 0.46 | 0.29 | 0.75 | 1.00 | 0.49 |
| Gross Ocean Product | 0.86 | 0.47 | 0.62 | 0.79 | 0.49 | 1.00 |

Table 2 showed there is a limit correlations between the ownership of marine fishing motor vessels, the rooms in star-related hotels and the berths for productive use. So it is reasonable to choose them as the input index for evaluating marine economy. On the other hand, the correlation coefficient between two types of output and inputs are significantly positive. Hence, when inputs are added, the desirable output "gross ocean product" and undesirable output "waste water discharged" is increased at the same time.

**Empirical Results**. Firstly, the economic efficiency (EE) was calculated by the CCR model in DEA for the 11 coastal provinces. It does not consider the environmental impact to marine economic, and makes the gross ocean product as output index, national ocean-related employed personnel, star-related hotel rooms by coastal regions and berths for productive use at above designed size seaports and the ownership of marine fishing motor vessels as input index, the results was shown in the "EE" column of Table 3. Then, using formula (1) on the undesirable outputs model (UOM), the environmental performance index (EPI) was evaluated with gross ocean product, and total volume of industrial waste water discharged as output indexes, and the inputs were same as before. Their results were shown in the "EPI" column of Table 3. In these two models calculation, the DEA solver PRO 9 software was used and constant returns to scale were assumed.

The value of EPI in Table 3 reflects the comprehensive efficiency. It evaluates marine economy of the coastal provinces and their efforts to protect the marine environment. It is almost consistent with the actual situation. In 2013 May, ocean monitoring department of China ocean bureau found in their monitoring for exceed discharge of sewage outfall: the highest is in Liaoning, its ratio of exceeding the standard is 100%, followed by Zhejiang, it is 71.4%, the rate of Fujian and Shandong were 44.4% and 25.7%[6]. In this paper, according to the EPI values evaluated by UOM model, the order from low to high is Zhejiang, Liaoning, Fujian and Shandong. It had virtually the same results with the actual investigation..

Table 3  Efficiency evaluation of marine economy for Chinese coastal provinces in 2011

| DMU | efficiency / Ranking | | Reduction rate of Fishing vessels (%) | | Reduction rate of Berths (%) | | Reduction rate of Number of Hotel Rooms (%) | | Reduction rate of Waste Water Discharged (%) | Reduction rate of Personnel (%) | Increasing rate of Gross Ocean Product (%) | Per Capita GDP (10000 yuan) |
|---|---|---|---|---|---|---|---|---|---|---|---|---|
| | EE | EPI | CCR | UOM | CCR(%) | UOM | CCR | UOM | UOM | UOM | CCR | |
| Liaoning | 0.49/6 | 0.22/7 | 91.2 | 98.8 | 0 | 57.7 | 0 | 69.8 | 79.2 | 48.4 | 103.8 | 5.07 |
| Hebei | 0.63/5 | 0.23/5 | 64.4 | 97.7 | 0 | 59.1 | 26.4 | 86.1 | 93.1 | 71.2 | 59.9 | 3.39 |
| Tianjin | 1.00/1 | 1.00/1 | 0 | 0 | 0 | 0 | 0 | 0 | 0 | 0 | 0 | 8.34 |
| Shandong | 1.00/1 | 1.00/1 | 0 | 0 | 0 | 0 | 0 | 0 | 0 | 0 | 0 | 4.71 |
| Jiangsu | 1.00/1 | 1.00/1 | 0 | 0 | 0 | 0 | 0 | 0 | 0 | 0 | 0 | 6.22 |
| Shanghai | 1.00/1 | 1.00/1 | 0 | 0 | 0 | 0 | 0 | 0 | 0 | 0 | 0 | 8.18 |
| Zhejiang | 0.42/9 | 0.16/9 | 95.2 | 97.9 | 0 | 83.0 | 0 | 80.4 | 86.0 | 46.5 | 139.9 | 5.92 |
| Fujian | 0.48/8 | 0.23/5 | 91.4 | 98.9 | 0 | 61.7 | 0 | 60.6 | 86.4 | 50.1 | 108.7 | 4.72 |
| Guangdon | 0.49/6 | 0.21/8 | 95.5 | 97.4 | 3.8 | 73.2 | 0 | 68.6 | 71.1 | 45.0 | 105.0 | 5.52 |
| Guangxi | 0.21/11 | 0.07/11 | 73.8 | 99.1 | 0 | 89.0 | 3.0 | 92.8 | 96.6 | 73.1 | 365.5 | 2.52 |
| Hainan | 0.24/10 | 0.13/10 | 89.4 | 99.6 | 0 | 70.8 | 0 | 88.6 | 46.1 | 75.5 | 321.7 | 2.88 |
| Mean | 0.63 | 0.48 | 54.6 | 62.7 | 0.35 | 45.0 | 2.7 | 49.8 | 50.8 | 37.3 | 109.5 | 5.22 |

Note: 1. Reduction rate of Personnel for CCR model and Increasing rate of Gross Ocean Product were not listed in the table, because all their values are 0.
2. The CCR model does not take into account the discharge of industrial waste water, so reduction rate of waste water discharged is only calculated by UOM model.

**The relation between EPI and EE.** It can be seen in the table 3, there is a strong correlation between the EPI and EE. Whether or not considering the environment benefit, the relative ranking of the two kinds of efficiency was almost invariant. According to their efficiency, all coastal provinces and municipalities in China can be divided into three levels. The first level was Shanghai, Jiangsu, Shandong and Tianjin, their two kinds of efficiency both were the highest, followed by Fujian, Hebei, Liaoning and Guangdong, and Zhejiang, Hainan and Guangxi had the lowest level. Further analysis found that, the relative difference of EE is smaller than EPI. If the first group is 100% for EE, second group is in 48% to 63%, and the minimum in the third group is more than 20%. But if considering the environmental benefits, the relative efficiency EPI varies greatly, the difference between EPI is bigger than between the EE. As the first group was 100% for EPI, the second was only in 21% to 23%, the third was below 20%. So for China's marine economy, the EE have close ties to the EPI. An area with the highest economic efficiency also has the highest environmental performance, and vice versa, low economic efficiency will lead to lower environmental performance.

**The relation between EPI and per capita GDP.** Economic efficiency is not equal to the level of economic development. If the degree of economic development in a regional is represented by per capita GDP, the results in this paper showed that the relations between EPI and the local economy were different in different stages. When per capita GDP is low (less than 30000 yuan), such as Hainan and Guangxi, the EPI is low; when per capita GDP is high (greater than 60000), the EPI is high too; but when per capita GDP is between them (about from 40000 to 60000 yuan), the EPI changes uncertainly, such as per capita GDP of Shandong was 4.71 million yuan in 2011, its EPI was 100%, but Zhejiang's per capita GDP wass close to 6 million yuan, the EPI was less than 17%. This shows that, in the development stage, effective environmental management is particularly important to improve the environmental performance.

**The relationship between EPI and industrial structure.** The calculation of two models show that, whether or not consider the environmental impact, to improve the economic efficiency under maintaining the current output, the reduction rate of marine fishing motor vessels is the highest, the average reduction rate in CCR model is 54.6%, and the average rate in UOM model is 62.7%. For considering the environmental impact, the UOM model reflect: the marine fishery had the lowest efficiency, and the marine tourism is the second. The average reduction rate of its corresponding evaluation index "rooms in star-related hotels" was 49.8%, marine transport industry is the highest efficiency in the three industries, the average reduction rate for its represent index "berths for productive use" was only 45%.

According the research, to improve the efficiency of marine industry, adjustment of marine fisheries is the common task for all non-effective area beside Tianjin, Shandong, Jiangsu and Shanghai. In addition, Liaoning, Hebei, Guangxi and Hainan should focus on improving the efficiency of tourism industry, Zhejiang, Fujian and Guangdong should pay attention to raise the efficiency of port and transport industry. At the same time, Hebei, Guangxi, Fujian and Zhejiang should carry out more stringent waste water treatment standards to improve the industries achieving more effective state, because their reduction rate of waste water discharged was the highest.